\documentclass[preprint,showpacs,preprintnumbers,amsmath,amssymb]{revtex4}
\usepackage{graphicx}% Include figure files
\usepackage{dcolumn}% Align table columns on decimal point
\usepackage{bm}% bold math
\begin{document}

\title{Dispersion of a single $4f$ impurity state in photoemission spectra of Yb/W(110)}

\author{Yu. S. Dedkov,\footnote{Corresponding author. E-mail: dedkov@physik.phy.tu-dresden.de} D. V. Vyalikh, M. Holder, M. Weser, S. L. Molodtsov, and C. Laubschat}
\affiliation{\mbox{Institut f\"ur Festk\"orperphysik, Technische Universit\"at Dresden, 01062 Dresden, Germany}}
\author{ Yu. Kucherenko}
\affiliation{\mbox{Institute for Metal Physics, National Academy of Sciences of Ukraine, 03142 Kiev, Ukraine}}
\author{M. Fonin} 
\affiliation{\mbox{Fachbereich Physik, Universit\"at Konstanz, 78457 Konstanz, Germany}}

\date{\today}

\begin{abstract}
Angle-resolved photoemission spectra of a monolayer of Yb on W(110) reveal energy splittings and dispersion of the Yb $4f$ states that are obviously due to their hybridization with W-derived valence bands. These effects occur at well defined points of the surface Brillouin zone although a smearing over reciprocal space is expected from the structural incoherence of the Yb and W lattices. We conclude therefore that dispersion is not related to the periodic arrangement of the $4f$ states but reflects the \textbf{k}-dependent interaction of a single Yb $4f$ impurity with W 
bands.      
\end{abstract}

\pacs{71.20.Eh, 73.20.-r, 79.60.-i}

\maketitle

In recent years, investigation of heavy-fermion  behavior in Ce and particularly Yb systems has attracted renewed interest in the framework of quantum phase transitions~\cite{Sachdev:2008,Gegenwart:2008}. As prominent examples, in  compounds of the type 
YbX$_2$Si$_2$ (X$=$Rh, Ir) the electronic properties may be tuned from superconductivity to magnetism by replacing the $5d$ element Ir by the isoelectronic $4d$ element Rh~\cite{Trovarelli:2000,Hossain:2005}. The compounds are, thus, thought to be on different sites of a quantum critical point where coexistence of these opposite properties may be expected at zero temperature. The low-temperature magnetic and conducting properties in both compounds are strongly influenced by effective masses of the conduction electrons that are by factor of thousand larger than the one of a free electron due to interactions of the 4$f$ states with valence bands. The effective masses relate to the band dispersions that can be studied most directly applying 
angle-resolved photoemission (PE).

Angle-resolved PE experiments reveal energy splittings and dispersion of the interacting $4f$ states as intuitively may be expected from a Kondo lattice~\cite{Danzebacher:2005,Danzebacher:2007,Vyalikh:2008}. However, these phenomena are not restricted to the region of the Kondo-resonance, where the PE final state is energetically degenerated with the $4f$ ground state, but are also observed for $4f$ emission at higher binding energies~\cite{Vyalikh:2006,Danzebacher:2006}. There, the fully screened PE final state is usually handled like a localized exciton~\cite{Johansson:1979} and energy dispersions are at first glance not expected. Interestingly, these phenomena occur around points in reciprocal space where the PE final state becomes energetically degenerated with dispersing valence states. Thus, the basic question arises whether the observed phenomena are necessarily related to a Kondo lattice, i.e. a periodic arrangement of  interacting $4f$ states or may also be expected for a single $4f$ impurity interacting with translational symmetric bands.

In order to answer this question, we performed angle-resolved PE experiments of an Yb monolayer on W(110). Divalent Yb grows epitaxially on the W substrate but its two-dimensional hexagonal lattice is incoherent to the rectangular centered substrate surface structure (see inset in Fig.~\ref{fig1}). As a result of this structural incoherence, interaction effects between 
substrate and overlayer are not expected to reveal an explicit wave-vector (\textbf{k} vector) dependence in the angle-resolved PE spectra but should be smeared in \textbf{k} space as a result of back-folding. Surprisingly, the $4f_{7/2}$ emission reveals strong energy splittings and dispersion around the $\overline{\Gamma}$ point of the surface Brillouin zone (SBZ), where no interactions with Yb-derived valence bands are expected but an energy degeneracy with underlying W bands takes place. From the fact that this phenomenon is not observed at other \textbf{k} points one can conclude that the effect is not related to an interaction of the incoherent lattices but reflects a \textbf{k}-dependent interactions of a single Yb impurity with the W bands. The phenomenon is properly described in the framework of the same simple approach to the periodic Anderson model (PAM) that previously had been successfully applied to the above mentioned YbX$_2$Si$_2$ heavy-fermion systems~\cite{Danzebacher:2005,Danzebacher:2007}. In this model the unhybridized $4f$ state is represented by a non-dispersive band at energy $\varepsilon$ that interacts with W derived bands as obtained from a regular band-structure calculation. In the 
present case the dispersionless $4f$ band is not caused by interatomic $f-f$ interactions, but reflects simply the effect of a Fourier-transform of a localized in real space $4f$ impurity into reciprocal space using basis valence-state wave functions of the substrate lattice. From our results we conclude, that the observed $4f$ dispersions in angle-resolved PE spectra are not 
necessarily the result of a periodic arrangement of $4f$ states but may be due to \textbf{k}-dependent interactions of a single $4f$ impurity with surrounding valence bands as well.
 
PE experiments were performed with a hemispherical electron energy analyzer using synchrotron radiation from the U49/2-PGM1 undulator beamline of BESSY (Berlin). The energy resolution was set to 100\,meV (FWHM) and an angular resolution better than 1$^\circ$ was used. Structurally ordered films of Yb were grown by vapor deposition at room temperature on a W(110) substrate. Prior to deposition the W surface was cleaned by repeated cycles of heating up to 1300$^\circ$\,C in an oxygen atmosphere at a pressure of 10$^{-8}$\,mbar and subsequent flashing up to 2300$^\circ$\,C. Low energy electron diffraction (LEED) patterns of the substrate reveal the sharp pseudo-hexagonal structure expected for the \textit{bcc}(110) surface while the LEED pattern of the Yb overlayer shows a hexagonal structure that is incoherent to the one of the substrate and characteristic for the \textit{fcc}(111) surface of Yb metal (inset in Fig.~\ref{fig1}). The base pressure in the vacuum chamber was below $5\times10^{-11}$\,mbar rising shortly to $1\times10^{-10}$\,mbar during Yb evaporation. No oxygen contaminations were found monitoring the O 2$p$ photoemission signal.

Yb films reveal strong layer-dependent shifts of the $4f$ emission~\cite{Martensson:1988} that may be used to calibrate the thickness of the film and allow for preparation of well defined monolayers. Figure~\ref{fig1} shows a series of angle-resolved PE spectra of the Yb/W(110) system taken for different Yb coverages at 110\,eV photon energy and normalized to the maximum of intensity. At this photon energy the W $5d$ emissions are close to a Cooper minimum of the PE cross-section~\cite{Yeh:1985,Molodtsov:2000} and the spectra are, therefore, dominated by Yb 4$f$-derived emissions. An emission angle of 5$^\circ$ off-normal was chosen in order to eliminate effects that appear close to normal emission and will be discussed later. At a coverage of 1 monolayer (ML) the PE spectrum consists of a single spin-orbit split doublet at 0.91\,eV binding energy (BE) superimposed by a flat step-like valence-band background. The spin-orbit splitting amounts to 1.28\,eV as characteristic for the $4f^{13}$ final state of divalent Yb metal~\cite{Schneider:1983}. Increasing the Yb coverage, the intensity of this doublet decreases and disappears at coverages exceeding 2\,ML. Instead, two new doublets with the same spin-orbit splitting arise at 0.54\,eV and 1.65\,eV BE, respectively. Signals at the same BE have been observed previously for a double layer Yb on the isoelectronic Mo(110) surface and were assigned to emissions from interface and surface atoms, respectively~\cite{Martensson:1988}. This assignment was supported by thermochemical model calculations, that may also be applied to the present case of the Yb/W(110) system since the relevant material parameters of Mo and W are almost the same. For coverages exceeding 2\,ML the interface component at 0.54\,eV BE becomes successively replaced by a new doublet at 1.21\,eV that is characteristic for the bulk emission of Yb metal~\cite{Schneider:1983,Martensson:1988}, while the BE of the surface component shifts at the same time  slightly from 1.65 to 1.73\,eV.

In the following we will concentrate on the monolayer Yb on W(110). Figure~\ref{fig2} shows a series of angle-resolved PE spectra taken at different emission angles along the $\overline{\Gamma}-\overline{S}$ direction (inset in Fig.~\ref{fig2}) of the SBZ of W(110) and 110\,eV photon energy. While at high emission angles only the $4f^{13}$ doublet is visible in the spectra, additional spectral features appear around normal emission that at first glance might be ascribed to residual valence-band emissions. On the other hand, the spectra can obviously not be explained by a simple superposition of $4f$ and valence-band contributions, since the $4f_{7/2}$ component becomes broadened and decreases in height with respect to the $4f_{5/2}$ emission.

This is illustrated in Fig.~\ref{fig3}(a) where spectra taken at emission angles of $0^\circ$ and $7^\circ$ are compared to each other together with the results of a least-squares fit analysis. In the framework of this analysis first the spectrum taken at $7^\circ$ was fitted by a superposition of two Lorentzians with the same line-width, and a $I(4f_{7/2})/I(4f_{5/2})$ intensity ratio of $1.15\pm0.02$ was obtained. Then, the normal emission spectrum was fitted using 4 Lorentzians with the same line-width as used above, replacing the former $4f_{7/2}$ line by components at 1.43\,eV, 1.04\,eV, and 0.76\,eV BE. Taking now the intensity ratio of the last two components with respect to the $4f_{5/2}$ line, almost the same value $1.17\pm0.02$ is obtained as for the $I(4f_{7/2})/I(4f_{5/2})$ intensity ratio of the $7^\circ$ spectrum. Thus, one might assume that these two components are derived from  the former $4f_{7/2}$ line and the third component at 1.43\,eV BE represents a valence-band emission. A possible candidate for the latter is the $5d$ band of the underlying W substrate that reveals particular strong emissions at 1.35\,eV BE around the $\overline{\Gamma}$ point and may also be responsible for the features close to the Fermi 
energy observed at higher emission angles.

Such an interpretation, however, is not fully satisfactorily since it does not explain the splitting of the $4f_{7/2}$ component. A similar splitting has been observed for the Ce "$4f^0$" signal of Ce/W(110) at a point in \textbf{k} space where the "$4f^0$" state becomes energy degenerate with the Ce-derived $sd$ band~\cite{Vyalikh:2006}. There, the splitting could be 
described in the light of the periodic Anderson model by formation of symmetric and antisymmetric linear combinations of wave functions of $4f$ and valence-band states. In the present case of the Yb/W(110) interface, band structure calculations show~\cite{Kucherenko:2008} that the bottom of $6s-$derived Yb valence band is located at around 4\,eV BE and only a crossing of the Yb $4f_{5/2}$ energy position is expected close the $\overline{S}$ point. There, however, no hints of a line splitting are observed what may be explained by the properties of the valence-band states, that are in case of Yb much more localized than in Ce and reveal almost Yb $6s$ character that does not hybridize with the Yb $4f$ states.
 
If interactions with the Yb valence band can be ruled out as a reason for the $4f$ splitting at the $\overline{\Gamma}$ point, there remain only interactions with the W $5d$ band as an explanation. Note, however, that at first glance from the incoherence of the lattices a smearing of respective effects over \textbf{k} space might be expected. Respective effects have not been observed for Ce/W(110)~\cite{Vyalikh:2006}, possibly because line-width and BE of the Ce "$4f^0$" state are much larger than for the Yb $4f_{7/2}$ component. In order to describe the interaction with the W band in spite of the lattice mismatch in the framework of the PAM one has to consider the Yb $4f^{13}$ hole state as an impurity that interacts particularly with W $5d$ states directed perpendicular outward to the W(110) plane. In order to determine the respective character of the W bands, band-structure calculations have been performed using the linear muffin-tin orbital (LMTO) method~\cite{Andersen:1975}. The results are shown by white dots in the Fig.~\ref{fig3}(b) where the size of the dots is a measure for the magnitude of the $5d$ admixture to the respective valence-band states. (Inset also shows experimental angle-resolved PE data for W(110) measured 
along the $\overline{\Gamma}-\overline{S}$ direction as a color plot.) Strong $5d$ admixtures to the W bands correspond to strong intensities in the PE signal as is expected from the larger cross section of the W $5d$ as compared to the W $6s$ states. The interaction with the $4f$ state in the light of PAM may now be handled in a similar way as for the Ce/W(110) system and the YbX$_2$Si$_2$ heavy-fermion compounds: The unhybridized hole state is considered as a dispersionless band at 
energy $\varepsilon=\varepsilon(\textbf{k})$ whereby the band character is in the present case not thought to be related to a periodic arrangement of $4f$ states but reflects simply a Fourier expansion of an localized impurity state in basis states of the W lattice. The hybridization matrix element is assumed to be proportional to the $d-$projected partial weight of the band states and described by the hybridization parameter, $\Delta$. Using the Anderson formalism in hole representation and excluding double-hole states ($4f^{12}$ configurations) from our consideration by setting the on-site Coulomb correlation energy $U_{ff}$ to infinity allows then for a diagonalization of the Hamiltonian. The obtained result is reminiscent of that of the single impurity Anderson model (SIAM). The important difference is, however, that now \textbf{k} dependence is achieved by replacing the density of states (DOS) used in SIAM by a \textbf{k}-resolved partial weights. Since $\varepsilon$ is given by the BE of the $4f_{7/2}$ component at \textbf{k} points far away from the $\overline{\Gamma}$ point the proportionality constant $\Delta$ (equals to 0.22\,eV in accordance with our earlier works on Yb systems) is the only adjustable parameter of this approach.
 
Figure~\ref{fig4} shows a series of calculated $4f$ spectra along the $\overline{\Gamma}-\overline{S}$ direction of the W(110) SBZ that reproduces all features observed in the respective experimental data. Note, that particularly the peak at 1.37\,eV BE in the normal emission spectrum should not be considered as a pure valence-band emission but represents a hybrid state with strong $4f$ admixture. Since in the calculations only the $4f$ character is shown its intensity relative to the unhybridized $4f_{5/2}$ component is weaker than in the experiment. Similar arguments hold for the features close to the Fermi energy that are caused by hybridization of the $4f_{7/2}$ state with an upwards dispersing band. Note, that also for the $4f_{5/2}$ state an intersection with the band is expected from dispersion. This takes place on halfway between the $\overline{\Gamma}$ and $\overline{S}$ points, and in fact the model calculations reveal there a change of the $4f_{5/2}$ line shape. This change, however, is  very weak because the $d$ admixture to the valence-band state is small and this effect is beyond the experimental resolution.

In summary, we have shown that the Yb $4f$ PE signals of Yb/W(110) reveal energy splittings and dispersions that are caused by hybridization of the Yb $4f$ states with W $5d$-derived bands. These effects appear at well defined points in reciprocal space although a smearing in \textbf{k} space is expected due to the incoherence of the lattices of Yb and W. Obviously the periodicity of the Yb lattice is not important for the effect, but the $4f$ states act as non-interacting impurity states and the dispersive properties observed in the PE experiment reflect the \textbf{k} dependence of the interaction. Applying this result to heavy-fermion systems one may conclude that the observation of $4f$ dispersions in angle-resolved PE experiments is not necessarily a proof for $f-f$ interactions within a Kondo lattice but may be caused by independent from each other Kondo impurities as well. 

This work was funded by the Deutsche Forschungsgemeinschaft, SFB 463, Projects B4 and B16. We would like to acknowledge BESSY staff for technical support during the experiments.

\end{document}